\documentclass[journal,12pt,3p]{elsarticle}

\usepackage{amssymb}
\usepackage{comment}
\usepackage{algorithmicx}
\usepackage[utf8]{inputenc}
\usepackage{algorithmicx}
\usepackage{algpseudocode}
\usepackage{algpseudocode}
\usepackage{multirow}
\usepackage{graphicx}
\usepackage{natbib}

\newtheorem{lemma}{Lemma}
\newcommand{\mod}{{\rm\ mod\ }}

\journal{Environmental Monitoring and Software}

\begin{document}
\bibliographystyle{plainnat}
\begin{frontmatter}

\title{Development of a data model to facilitate rapid Watershed Delineation}
\tnotetext[label0]{This is only an example}

\author[label1]{Scott Haag}
\address[label1]{Academy of Natural Sciences, Drexel University 1900 Benjamin Franklin Parkway, Philadelphia, PA 19103, USA}

%%\cortext[cor1]{I am corresponding author}
%%\fntext[label3]{I also want to inform about\ldots}
%%\fntext[label4]{Small city}

\ead{smh362@drexel.edu}
 
\author[label2]{Ali Shokoufandeh}
\address[label2]{Department of Computer Science, Drexel University, Philadelphia, PA 19105, USA\fnref{label4}}
\ead{ashokouf@cs.drexel.edu}

\begin{abstract}
An efficient model to store and retrieve surface watershed boundaries using graph theoretic approaches is proposed. Our approach utilizes three algorithms and accepts as input standard digital elevation models derived stream catchment boundaries. The first is called Modified Nested Set (MNS), which is a generalized depth first graph traversal algorithm that searches across stream reaches (vertices) and stream junctions (edges) labelling vertices by their discovery time, finish time, and distance from the root. The second is called Log Reduced Graphs (LRG), which creates a set of logarithmically reduced graphs from the original data, to store the watershed boundaries. The final algorithm is called Stitching Watershed, which provides a technique to merge watershed boundaries across the set of graphs created in the LRG algorithm. This technique is show to provide significant advantages in processing, storage, and retrieval complexity when compared to hypothetical data models.

\end{abstract}


\end{frontmatter}

% \linenumbers

\section{Software and/or data availability}
This paper is focused on a novel data model to store and access watershed boundaries, where the outputs are the algorithms.  The proposed algorithms were applied to example datasets within the Delaware River Watershed using currently available data for the United States of America. Worldwide datasets such as the Shuttle Radar Topography SRTM and associated products such as Hydrological data and maps based on SHuttle Elevation Derivatives at multiple Scales (HYDROSHEDS) [1] would allow these methods to be applied in a consistent fashion world-wide.  

Existing applications that derive watershed boundaries take a number of forms including graphical outputs, modelling programs, and software as a service.  We believe that most of these approaches could benefit from our proposed data structure and therefore we avoided creating software specific results.  If these methods are adopted it will be incumbent on the developers to implement them within their unique environments.  While we used specific software and code to implement the method, we do not at this time offer them for download, as they would provide limited utility outside of our use case.
 
\section{Introduction}

 Any model of natural processes requires an understanding of the connection between the physical properties of mass, energy, and momentum.  These analyses are fundamental to simulating how perturbations might impact natural systems.  In this paper we focus on the hydrodynamic connections caused by water flowing over terrestrial surfaces and through river and stream networks.  This type of modelling is collectively know as "watershed modelling" and is fundamental to a variety of disciplines including planning for urban and natural landscapes, measuring ecological connectivity, tracking contaminates sources, and modeling stormwater surge and flooding.  Specifically, we focus on how to represent vectors of connected streams and associated physical attributes within a series of graphs that are reduced in complexity to store and retrieve continuous watershed boundaries to limiting both search and storage costs.  We compare our approach to hypothetical techniques to store watershed boundaries across a topological surface using examples from the $\approx$ 30,600 km$^2$ Delaware River Watershed (Figure 1). Source data for our study area is a the nationally available Watershed Boundary Data set (WBD) the National Hydrography Plus Version Two created by Horizon Systems and the United States Environmental Protection Agency EPA (NHDPlusv2) [2].
 
 A key feature within any mechanistic watershed model is how stream topology is stored and processed (the data model).  The `best' data model for a specific application can significantly reduce its amortized computational complexity and processing time [3].  Depending on the level of documentation the exact data model employed within a particular watershed modelling tool can be difficult to determine. Stream topology is typically stored as a sparse matrix as described in the Routing Application For Parallel Computation of Discharge (RAPID) [4] or as an adjacency list as in the Soil and Water Assessment Tool (SWAT) [5].  Liu and Weller 2008 [6] use a binary search tree to store and their adjacency list.  Adjacency lists require a recursive search through a steam topology list, while the sparse matrix approach allows a single database sweep but it is more costly in data storage terms vs the adjacency list.  We provide a general alternative to the adjacency list and the sparse matrix representation called the Modified Nested Set (MNS) index that provides two primary advantages over the aforementioned techniques; 1) watershed boundaries and downstream flow lines can be created without recursively searching the database or storing every stream to stream relationship, 2) boundaries can be created by returning aggregated versions of upstream areas. Which means that less objects are merged to create the final watershed boundary.  Additionally, we provide an approach to store the polygonal boundary that represents a stream's watershed using two algorithms Log Reduced Graph (LRG) and Stitching Watershed (SW) respectively.  The LRG algorithm is designed to store increasingly reduced versions of the original watershed boundary by merging adjacent stream reaches.  The SW algorithm takes the output from the LRG algorithm and merges vertexes to create a watershed boundary for any stream reach.  We compare results against two example data models, the first called baseline model assumes that polygons representing lateral watershed are merged individually on demand for any selected $v(s)$, the second called the processed model assumes that all watershed boundaries are calculated and stored for every vertex $v(s)$ in graph $G$.  We show that our data model requires significantly less data storage vs the processed data model and significantly fewer objects and polygonal nodes to merge in the query vs the baseline data model.

In this paper we focus on watershed modelling as a function of watersheds delineated by land surface elevation.  We recognize that this is only a part of a comprehensive watershed modelling effort (see Daniel et al. [7],  and  Singh and Woolhiser 2002 [8]).

The approach we employ here requires a simplification of riverine networks by assuming that water drains in only one direction at all times.  This can be violated in specific instances such as tidal areas, islands in the main river stem and river divergences (e.g. Oxbow). In our study area 98.8$\%$ of stream reaches are considered the main stem 1.2$\%$  are considered the minor reach which could cause divergent flow, this could be a larger issue in other watersheds.  Although this is a potential shortfall of the nested set model, modelling river flow with only one possible downstream edge is a common approach in hydrological modelling and is applied in several popular models, such as the Storm Water Management Tool (SWMMT) ( http://support.chiwater.com/support/solutions/articles/38406-flow-routing ) ‘RAPID	can	currently	only	accommodates one	unique	 downstream	reach	for	each	river	reach. [4].  To address this issue we assume that flow is 100\% within the major divergence and minor divergence are indexed by assuming no upstream connectivity, but can still be traversed on the downstream junction.  This allows downstream accumulations to be correct for all values with potential inaccuracies located only inside of the minor divergence. It can be shown that this can be fixed by applying a new discovery $d$ time value to a minor divergence for search purposes only.  

\section{Material and Methods}
\subsection{Notation}
\label{subsec1}

Before we can discuss the specific methodological procedures that we apply to this problem we need to define some of the  notation that will be used throughout the paper.  

This paper focuses on methods to embed a representation of a watershed $\Omega$ stored as a series of streams connected at junctions denoted by $S$ and $J$ respectively. We create an abstract representation of this entity in terms of a graph $G(V,E)$ with a vertex $v = v(s)$ for every stream $s={(s_1,s_2,...,s_k)}$ and an edge for every stream junction $e = e(j)$ defined as the intersection of $k$ incoming streams $s_1 , ... , s_k$ and one outgoing stream $s_o$. Every $e$ can be defined by the $e_{[s_1,s_2,...s_k;s_o]}$ (Figure 2 and 3).  Figure 2 shows an example of several stream reaches $s$ and junctions $j$ inside of the study area, Figure 3 shows the same stream reaches relabelled in terms of $G(V,E)$.

We define several attributes for $G$ including the $root(G)$ as the only vertex in G without any downstream $e$, $leaf(G)$ as a vertex that has no upstream $e$, and $\lambda(G)$ as the height of $G$ defined as the maximal number of $v$ that a path traverses between the $root(G)$ and any $leaf(G)$.

Additionally, we describe several attributes for every $v$ including the local watershed catchment $\omega(v)$ which describes the area where water flows into $v$ due to over land and groundwater (non-fluvial) flow. This is also described in the literature as the lateral flow for $s_k$ and is stored within a Geographic Information Systems as a series of connected $x$ and $y$ vertices, the total watershed or $\Omega(v_k)$ described as the union of all local catchments $\omega(v)$ that are upstream of vertex $v_k$ including the starting vertex $\omega(v_k)$ itself. 

To create $\Omega(v)$ we apply three algorithms; the first is called the Modified Nested Set algorithm that for all $v in V$ assigns labels $d(v)$ (discovery time), $f(v)$ (finish time), and $\delta(v)$ (distance from the root). The second algorithm called Log Reducing Graphs is used to create a set $S$ of log reduced Graphs $S = \{ H_{0}(U_0,R_0),...,H_l(U_l,R_l)\} $, a local watershed boundary $\omega(u)$ is also created for all vertexes $u$ in $U$ of graphs $H$ in set $S$.  The third algorithm called Stitching Watershed Algorithm identifies and unions catchments from vertexes $\omega(u)$ from the set $S$ to create a watershed boundary $\Omega$ for vertex $v$.

Their are two disparate uses of the term vertex within this paper, the first common in the Geographic Information Systems (GIS) community describes vertexes as the \textbf{x} and \textbf{y} coordinates pairs used to define vector objects (points, lines and polygons); while in graph theory where vertexes describe nodes connected by edges.  For the rest of the paper we differentiate between the two calling vertexes in GIS  nodes leaving in place existing definitions of vertexes in graph theoretic terminology.

\subsection{Source Data}
\label{subsec2}
A Geospatial version of the NHDPlusV2 files was downloaded from http://www.horizon-systems.com/nhdplus/NHDplusV2\_data.php for the Delaware River Watershed and stored inside a PostGRES Relational Database Management System (RDMS) version 9.1.12 running on a suse linux server running the PostGIS suite of geospatial functions version 2.1.4 r12966. The data was transferred into PostGres using the PostGIS shapefile and database file loader. Individual stream reaches were stored in a table with multiple attributes including a representation of the stream center-line stored as a geometry LINESTRING, a representation of each reaches drainage area stored as a geometry POLYGON representation, a unique integer identifier, a designation of flow order, and a list of upstream and adjacent stream reaches.  Spatial data was projected into the Universal Transverse Mercator, (UTM) Zone 18 North, meters World Geodetic Survey 1984 (WGS84), Spatial Reference Identifier (SRID) 32618 using the embedded PostGIS function ST\_Transform.   Projecting the spatial data allows PostGIS take take advantage of the spatial geometry functions that are part of the POSTGIS package, the original data is in a geographic projection system North American Datum of 1983 (NAD83) SRID 4269.  

While the data was stored in PostGres, algorithms MNS was implemented in the R program for statistical computing.  Algorithm LRG and SW was implemented within PostGres as a function withing a Standard Query Language (SQL) query. The R package RODBC [9] was used to transfer data between R and PostGres.

\subsection{Modified Nested Set Algorithm}
\label{subsec3}
The NHDplusv2 database defines for each reach the terminal reach identification or the unique hydrological sequence of the last reach above the confluence of an estuary, bay or oceanic water body as the TerminalPa (Teminal Path) variable [2]. We treated the data as a tree (a uncycled directed graph) and ran a Modified Nested Set (MNS) Algorithm (Algorithm 1) starting with the root node (hydro sequence 200005438) and all stream reaches where the TerminalPa = 200005438 or the confluence of the Delaware River with the Delaware Bay geographic location -75.35717 , 39.27537. This resulted in a file containing  $|V(S)|$ = 14,718 reaches covering a watershed 30,600 km2 (Figure 1). As the MNS algorithm crosses over $G$ starting from the root it records the discovery time $d(v)$, finish time $f(v)$, and distance from root $\delta(v)$ for each $v_k$.  The $d(v)$ and $f(v)$ values are equivalent to the nested set algorithm described by Celko [10].  The $d(v)$ value is an incremental value that is unique for each $v$ between 1 and $K$ as the algorithm moves up the tree to find new vertexes. No $d(v)$ value is less then than a downstream $d(v)$ value and no two reaches share a $d(v)$ value, while the $f(v)$ value calculates the highest $d(v)$ value that is upstream of a reach,  multiple reaches might share the same $f(v)$ value.  Figure 4. shows the results of the MNS algorithm on the example watershed used in Figures 2 and 3, stream reaches are labeled with three attributes ($d,f,\delta(v)$) from left to right respectively.   Leaves on $G$ can be identified where the $d(v_k)= f(v_k) $. The MNS algorithm runs in exactly 2$k$-1 where $k$ = the numbers of vertexes $|V|$ in graph $G$. Each edge $e$ is crossed twice (once up and once down) and each $v$ has only one downstream edge $e$ (except for the root which has none). A unique key using a B -tree Index was  created on $d(v)$ to guarantee individual vertex retrieval times in $\log{|V|}$ and to ensure cardinality of the result set. The MNS algorithm is a modified depth first (DF) graph search algorithm and runs in linear time based on the number of vertexes $|V|$ in graph $G$.
\newpage{}
\textbf{Algorithm 1. Modified Nested Set} 
\begin{footnotesize}
\begin{algorithmic}[1]
\State {\bf Input:} Graph $G=(V,E)$.
\State {\bf Output:} $\forall v\in V$: 
\State \ \ \ \ \ \ \ \ \ \ \ \ \ the discovery time $d(v)$, 
\State \ \ \ \ \ \ \ \ \ \ \ \ \ finish time $f(v)$, 
\State \ \ \ \ \ \ \ \ \ \ \ \ \ distance from root $\delta(v)$.  
\For {$v \in V$} do
   \State $d(v) \leftarrow$ {\sc null} 
   \State $f(v) \leftarrow$ {\sc null} 
   \State $\delta(v) \leftarrow$ {\sc null} 
 \EndFor
\State $\tau \leftarrow 1$  \Comment Universal clock
\State $\Delta \leftarrow 0$ \Comment Current distance from root
\State MNS($G,root$) \Comment {Initiate Traversal}
\Statex
\Procedure{MNS}{$G,v$}
\If{$d(v)$ == {\sc null}} \Comment Vertex is not discovered 
    \State $d(v) \leftarrow \tau$ \Comment Set discovery time
    \State $\delta(v) \leftarrow \Delta$   \Comment Set distance from root
    \State $\tau \leftarrow \tau + 1$ \Comment Set current discovery time
\EndIf
\For {$w \in V$ with $(w,v) \in E$} 
    \State $\Delta \leftarrow \Delta+1$  \Comment Set current distance from root 
    \State MNS($G,w$) \Comment{ Move up the tree to the first vertex in set $w$}
    \State $\Delta \leftarrow \Delta - 1$ \Comment Set current distance from root
\EndFor
\State $f(v) \leftarrow \tau$ \Comment {Set finish time}
\If {$v==root$} \Comment If current distance is below root
        
\State return() \Comment end procedure
\EndIf
\EndProcedure
%\Statex 
\end{algorithmic} 
\end{footnotesize}
 
\subsection{Log Reduced Graphs Algorithm}
\label{subsec4}

For each $v$ the total watershed boundary $\Omega(v_k)$ is the merged geospatial boundary of all upstream catchments $\omega(v)$, in the example where $d(v)$ = 1 (the root of graph $G$) there are 14,718 vertex boundaries $\omega(v)$ merged to create the watershed boundary for the root vertex $\Omega(root)$.  A catchment is within another catchment's watershed when the $d(v)$ value is between the reaches $d(v)$ and $f(v)$ value as described in Joe Celko's Trees and Hierarchies in SQL for Smarties [10].  In SQL this can be accomplished using a self join on the table storing the results of the MNS algorithm (Appendix 1).  The polygon that describes the watershed boundaries for each reach can be created using the included PostGIS spatial function ST$\_$Union (Appendix 1). 

We applied a second algorithm called Log Reduced Graphs to $G$ with the result being a set of graphs denoted as $S$.  To accomplish this we selected a base value denoted as $b$, and selected all $v$ in the original graph $G$ where $\delta(v)$ is a proper multiple of $b$. If $b$ = 2 and $\delta$ = \{0,1,2,...,16\} any vertex with values $\delta$ in \{0,2,4,...,16\} would be selected for the reduced graph vertexes conversely vertexes with $\delta$ values in \{1,3,5,...,15\} would be unioned with their closest down stream vertexes that is a proper multiple of $b$. The LRG algorithm then creates a reduced graph by combining all the local watershed boundaries $\omega(v)$ that are upstream with a value of $\delta$ but is downstream of the next factor of $b$ (Figure 6).  This new vertex is saved within graph $H$ until all vertexes $v$ in graph $G$ are processed (for a graphical example refer to Figure 5). The LRG algorithm is called recursively, increasing the reduction level $rf$ by 1 for each iteration and substituting graph $H$ for $G$, this continues until the last graph in set $S$ contains one vertex the root.  Because $rf$ is set to start at $0$ the first graph in set $S$ will be the original graph $G$.\\ \\
\textbf{Algorithm 2. Log Reduced Graphs Algorithm} (LRGs)\label{LRG} 
\begin{footnotesize}
\begin{algorithmic} [1]
\State {\bf Input:} 
\State Graph $G(V,E)$ with $V$ labelled from algorithm MNS $\{d(v)$, $f(v)$, and $\delta(v)\}$.
\State $\omega(v)$ \Comment Spatial Catchment Boundary for each Vertex in $G$
\State $b$ \Comment A constant $>=$ 2 
\State {\bf Output:} 
\State Set of log reduced Graphs $S = \{ H_{0}(U_0,R_0),...,H_l(U_l,R_l)\} $.
\State$\forall n\in N$: Lateral Drainage Boundary (Polygon) for each Vertex $\omega(n)$.
\State $rf \leftarrow 0$          \Comment Initiate the reduction depth
\Procedure{LRGs}{$G,b,rf=0$} \Comment Initial call to formation of log reduced set
%%\For {$b = 0$, $b{+}{+}$, while $b \preceq $ log $ (\Delta(g)$) + 1  } 

\State $H$=($U$=$\emptyset$,$R$=$\emptyset$)    \Comment Create Empty graph $H$ with Vertices $U$ and Edges $R$
 
    \For {$v \in V$} 
        \If {$\delta(v)$  mod  $b^{rf} = 0$} \Comment If vertex $\delta$ is a factor of $b^{rf}$
            \State $U\leftarrow U \cup \{v\}$ \Comment Copy the Vertex from $G$ to $H$ including labels from the SW algorithm
 
        \EndIf
    \EndFor
 
    \For {$u \in U$} 
        \For {$v \in V$}    
            \If {$d(u) \geq d(v)$ and $f(u) \leq d(v)$ and $\delta(u) =  \delta(v) + b^{rf} $ }
                \State $R  \leftarrow  R  \cup (u,v) $ \Comment Set Edges for $H$
            \EndIf
            
            \If {$d(u) \geq d(v)$ and $f(u) \leq d(v)$ and $\delta(u) < \delta(v) + b^{rf} $ }
                            \State $\omega(u) \leftarrow \omega(u) \cup \omega(v)$  \Comment Union Spatial Boundaries
            \EndIf
         \EndFor
    \EndFor
\State $S \leftarrow S \cup H(U,R)$ \Comment remove Vertex's that exist on more the on graph $H$ and save to set $S$

\If { $|U| >$ 1 }  \Comment Call LRG until only the root remains
\State $rf \leftarrow rf + 1 $ \Comment Set reduction depth
\State {LRG(}{$H,b,rf)$} \Comment Call log reduced set with increased depth
\EndIf
 
\State $rf \leftarrow rf + 1$  
\While {$rf < |S|$}
\For{$v \in H_{rf}(U,R)$}
\If {$\delta(v)$  mod  $b^{rf + 1 } = 0$} 
\State     $U\leftarrow U\setminus\{v\}$
\For{$(u,v) \in R$} 
\State     $R\leftarrow R\setminus\{(u,v)\}$
\EndFor
\EndIf
\EndFor
\State $rf \leftarrow rf + 1$
\EndWhile
\EndProcedure
\end{algorithmic}
\end{footnotesize}
Lemma 1 shows that the number of graphs returned from the LRG algorithm is bounded by $\left \lceil \log_b(\lambda(G))\right \rceil $.  For example using $b=2$ and a max($\delta$) = 600 for $G$ this algorithm would run recursively 10 times (i.e., $2^9 < 600 <  2^{10}$) and creates reduced $S=\{H_0,H_1,...H_{10}\}$. As a final step the LRG algorithm removes the set of vertexes and edges that exist inside of multiple graphs in set $S$.  Therefore any individual graph $H$ in set $S$ could be an unconnected graph, to create a final watershed boundary we need one last algorithm that moves across the reduced graphs $H$ in set $S$ connecting them and stitching together a watershed boundary.

\begin{lemma}
Given a directed graph $G(V,E)$ of maximum height $\lambda(G)$, let $S=\{G_0,...,G_{|S|}\}$ denote the set of reduced graphs obtained from LRGs algorithm with base $b\ge 2$, then $|S| \le \log_b(\lambda(G))$.
\end{lemma}
{\bf Proof.} We will first examine the recursive calls that generate consecutive reduced graphs, i.e., $G_{rf+1}\leftarrow  {\rm LRGs}(G_k,b,rf+1)$. Let $\lambda(G_{rf})$ and $\lambda(G_{rf+1})$ denote the length of longest paths in graphs $G_{k}$ and $G_{rf+1}$, respectively. 

Without loss of generality, assume $\left <u_1,...,u_{\lambda(G_{rf})}\right >$ denotes the path of maximum length in graph $G_{rf}$. The LRGs algorithm will contract this path by forming groups $$\{u_1,...,u_b\}, \{u_{b+1},...,u_{2b}\},...,\{u_{(\ell-1)b+1},...,u_{\ell b}\},$$
for $\ell = \left\lceil {\lambda(G_{rf}) \over b}\right \rceil$.  These  groups will in turn be replaced by vertices $\{w_1,..., w_\ell\}$ in graph $G_{rf+1}$, where $w_i\leftarrow\{u_{(i-1)b+1},..., u_{ib}\}$, $1\le i\le \ell$. This implies that $\lambda(G_{rf+1}) \le \lambda(G_{rf}) / b$. An inductive backward analysis shows that $\lambda(G_{rf+1}) \le \lambda(G) / b^{rf}$.  We note that for the final graph $G_{|S|}$ in $S$,  $\lambda(G_{|S|}) = 1$. This implies that $\left \lceil {\lambda(G)\over b^{|S|}} \right \rceil = 1$, which in turn implies $b^{|S|}\le \lambda(G)+1$. Solving for the size of $S$ we have $|S|\le \log_b (\lambda(G))$.

\begin{lemma}
Given a directed graph $G(V,E)$ of maximum height $\lambda(G)$ and the  contraction constant $b\ge 2$,  the computational complexity of the $LRG(G)$ is $O(|V|)$.
\end{lemma}
{\bf Proof.} Similar to the notations used in Lemma 1, let ${\cal P} = \left <u_1,...,u_{\ell b}\right >$ denote the vertices along a typical path processed by LRG that results in contracted path  $ {\cal P'} = \left <w_1,w_2,...,w_{\ell}\right >$. Note that, each $w_i\in {\cal P'}$ corresponds to $b$ vertices since $|{\cal P'} \le |{\cal P}| /b$. Extending this argument to each reduction step implies $|V_{\ell+1}|\le |V_{\ell}| /b\le {|V|/ b^\ell}$. As a result, the overall complexity of all recursive calls for the LRG algorithm can be bounded by $$\sum \limits_{\ell=0}^{|S|} |V_\ell|\le \sum \limits_{l=0}^{|S|} ({V /b^l})=|V|\left ( {1 \over b^0} + {1 \over b^1} + ... + {1 \over b^l}\right ).$$ Using the fact that $b\ge 2$, the computation complexity of the LRG algorithm can be bounded above by $O(|V|)$. 
 
\subsection{Creation of watershed boundaries (Watershed Stitching Algorithm)}
\label{subsec5}

To assemble the boundary for any stream reach $v(s_k)$ in the original graph $G$ a query is used to cross the set $S$ of graphs $H$ , identifying and unioning local vertexes $u$ watershed boundaries $\omega(u) \in H$ (Figure 5 and 6).  This algorithm is essentially a depth first graph traversal algorithm, except that because individual graphs $H$ in set $S$ may not be fully connected it must also search across multiple reduced graphs in set $S$.  The first step is to identify the graph to be searched, next is the distance from the root to search in each graph, and last to identify vertexes that are upstream of original search vertex $v(s_k)$.

To identify the correct graph to search in set $S$ the first step is to determine $r\leq |S|$, the highest power $rf$ of  $b$ for which the distance from the root is proper multiple of $b^rf$.  For example the root of $G$ always has a distance from the root of $\Delta(v) = 0$, therefore it always searches the last graph in set $S$ or $H_{|S|}$ regardless of the value of $b$. As a second example take the distance from the root value of $\Delta(v_k)= 6$, where the reduction factor $b= 2$, and set $|S|= 10$.  The SW algorithm identifies graph $H_1$ as the correct search level because $6 \mod 2 = 0 $ and $6$ mod $2^2 \ne 0$. Following this step, SW uses the nested set data structure to search within $H_{rf}$ identifying all vertexes that are upstream of the search vertex (including the vertex itself) whose distance from the root $\Delta(v)$ is less then $b^{l+1}$.  Once identified vertexes are merged into the global watershed boundary variable $\Omega(v)$. Lastly, SW identifies all vertexes $u$ in graphs $H_{rf}$ that are upstream again using the nested set algorithm and where the distance from the root exactly equals $\Delta$ = $b^{rf+1}$.  The SW algorithm is then recursively called, this time replacing the original search vertex with the identified upstream vertexes $u$. These vertexes will be found somewhere between the $H_{l+1},...,H_{|S|}$ level graphs.  In the first example where vertex $v$ was the root of $G$ and $\Delta(v)$ = 0 because it searched the last $rf$ of $S$ no vertexes will be found above $v$ and therefore the SW algorithm will stop after 1 iteration.
\\
\\
\textbf{Algorithm 3. Stitching Watershed Algorithm}
\begin{footnotesize}
\begin{algorithmic}[1]
\State {\bf Input:}  $S = \{ H_{0}(U_0,R_0),...,H_l(U_l,R_l)\} $ 
\Comment Set of Log Reduced Graphs $H$ with vertices $U$ and Edges $R$ 
\State{} \Comment and labels from LRG Algorithm
\State \ \ \ \ \ \ \ \ \ \ \  $\omega(u)$, \Comment Vertex catchment
\State \ \ \ \ \ \ \ \ \ \ \  $v$, \Comment Vertex From $G$ 
\State \ \ \ \ \ \ \ \ \ \ \  $\delta(v)$, \Comment Vertex distance from Root in $G$
\State \ \ \ \ \ \ \ \ \ \ \  $b$ \Comment Base value used in LRG algorithm
\State {\bf Output:} $\Omega_v$ \Comment Geometric Watershed Boundary for Vertex $v$
\State  $\Omega_v \leftarrow$ null \Comment Set the Watershed Boundary to null
\State{SWA($S,v,b$)\Comment {Initiate Stitching Watershed Algorithm}}
\Procedure{SWA}{$S,v,b$}
\State $rf^*$ = argmax$_{rf \in \{0,..., |S|-1\}}$ $\{\delta(v)$ mod $b^{rf} \neq$ 0\}
  \For {$u \in U_{rf^*} $} 
    \If {$d(u) = d(v)$}
    \State $\Omega_v  \leftarrow  \Omega_v \cup \omega(u)$ 
    \Comment Append to the output
    \EndIf
    \If {$d(w) \geq d(v)$ and $d(w) \leq f(v)$ and $\delta(w) = \delta(v) + b^{rf^*}$}
        \State SWA($S$,$u$,$b$,$\Omega$) \Comment Recursively call SWA
    \EndIf
\EndFor
\EndProcedure
\end{algorithmic} 
\end{footnotesize}

\begin{lemma}
Given a directed tree $G(V,E)$ of maximum height $\lambda(G)$ with an average graph bandwidth of $\nu=\nu(G)$, a constant contraction parameter $b\ge 2$, $S$, a set $S$ of reduced graphs returned generated by LRG algorithm, and labelled by MNS algorithm then SW runs in $O(\nu (b+\nu \log_b (n/\nu)))$.
\end{lemma}
%{\bf Proof for SW runs in $O$($\nu\log_b(|V|)$.} 
{\bf proof.}\\
We first note that for graph with $n=|V|$ vertexes and bandwidth $\nu$ and contraction constant $b\ge 2$, vertices of $G$ are expected to belong in one of layers $b^0,b^1,...,b^\ell$, with $\ell=O(\log_b (n/\nu))$, which follows from the fact that $b\ell=n/\nu$. We will use $T_{v,r}$ to denote the sub-tree rooted at node $v$ entirely between layers $b^r$ to $b^{r}+b$. Next, we estimate the complexity of recursive calls made by SW for computing watershed boundary $\Omega_v$ of a vertex $v\in V$. We will denote this complexity by $C_v$. We will consider two distinct case, $\delta(v)=b^r$ for $0\le r\le \ell$, or $b^{r-1}< \delta(v)<b^{r}$ for $1\le r\le \ell$.
\\
Assume $\delta(v)=b^r$ for $0\le r\le \ell$, then it is not hard to see that the complexity of computing $\Omega_v$ is the union of watershed induced by $T_{v,r}$ (denoted by $\Omega_{T_{v,r}}$) and the $\Omega_u$ over all vertices $u$ that are leaves of $T_{v,r}$ and the roots of new sub-tree $T_{u,r+1}$. First, observe that there exists a largest $k$ such that $v$ is the root of a sub-tree in $G_k\in S$ for which the value of $\Omega_{T_{v,r}}$ is already computed in LRG algorithm. We observe that the expected number of leaves for $T_{v,r}$ is $\nu$, and as a result we can bound the complexity $C_v$ for computing $\Omega_v$ using the recurrence:
$$C_v= (1+\nu(\ell-r+1)),\ {\rm for\ } 0\le r\le \ell-1.$$ 
Using the fact that $\ell=O(\log_b (n/\nu))$ we get the bound $C_r= O(1+\nu \log_b (n/\nu))$. 
\\
Next, we will consider the case $b^{r-1}< \delta(v)<b^{r}$, for $1\le r\le \ell$. Here, the complexity $C_v$ consists of two parts: the cost of computing $\Omega_{T_{v,r}}$ and the sum of the terms $\Omega_u$ over all vertices $u$ that are the leaves of $T_{v,r}$. Similar to previous case, the cost of computing the watershed of all leaves $u$ for $T_{v,r}$ which are the roots of  $T_{u, r+1}$ is  $O(\nu \log_b (n/\nu))$. In contrast, since $\delta(v)\neq b^{r}$ for any $0\le r\le \ell$ we can not find the a sub-tree in $G_k\in S$ which has node $v$ as a root. As a result, the complexity of computing $\Omega_{T_{v,r}}$ is proportional to number of nodes in $T_{v,r}$. Since the number of layers and bandwidth of  $T_{v,r}$ are at most $b$ and $\nu$, respectively, the bandwidth  the complexity of computing $\Omega_{T_{v,r}}$ can be bounded by $\nu b$. This implies $$C_v= \nu (b+\nu \log_b (n/\nu)).$$ 
Combining the two cases we have $C_v= O(\nu (b+\nu \log_b (n/\nu))).$

\section{Results}

\subsection{Algorithm Performance Metrics}
\label{subsec5}
We compared our proposed methodology in three dimensions which corresponded to the outputs of the three algorithms discussed here.  Next we compare computational costs in terms of preprocessing, storage, and query complexity of the LRG and SW algorithms to hypothetical data models.  We note that the baseline model and the processed model require the same number of calculations, but they differ in use cases where storage vs query complexity are issues. 

The first is the size of the index necessary to store the connections between each vertex $v$.  The two standard methods are the sparse matrix as described in River Network Routing on the NHDPlus Dataset [4] and the adjacency array as described by A Geodata Model and GIS Interface for Swat [5]. The MNS algorithm as described in this paper requires two attributes to be stored for any vertex $v$, the discovery time $d(v)$ and the finish  time $f(v)$.  The MNS index therefore will grow in constant space as the number of vertexes $v$ increases.  The Sparse matrix approach stores a Boolean value for every catchment to catchment relationship and therefore has space complexity $\Omega(v^2)$. The adjacency list approach stores directly adjacent vertexes $v$ and is equivalent in size to the number of edges $|E|$ of the graph. Because each vertex $v$ can only have one downstream edge the total number of $|E|$ = $|V|$-1 (the root does not have any downstream edges).  Therefore adjacency list will also grow in constant time as the number of vertexes  increases.  The advantage of the matrix and MNS approach is that each upstream vertex $v$ can be found in a single database sweep vs adjacency list which requires a recursive search through the database.  In addition some matrix representations only story upstream adjacent reaches not all upstream reaches, therefore recursive searches through the matrix would be required in a similar manner as the adjacency list. A disadvantage of the MNS index is the complexity to update the index given new information is considerable, for example if a new root is added to graph $G$ it would require updates to all index values for the entire graph.  While  problematic in this use case the temporal frequency of remotely sensed elevation data and corresponding edits to the flow direction graph is much smaller then the number of queries making it a relatively minor issue as data is collected in increasing temporal frequency this could become an interesting research focus.

\subsection{Data model preprocessing cost}
\label{subsec5}

The baseline data model in our example has no processing costs because we are not accounting here for the original creation of the DEM and the flow direction grid and all three data models require the same source data-sets.  The processed data model requires the union of $~$ 2.31 m polygons and $~$ 360m nodes.  The proposed model requires between $~$ 29.2 - 18.8 thousand polygons and  7.3 - 3.8 m nodes depending on the value of $b$ that is selected.  This is a reduction of between 98-99\% of the preprocessing costs.
 
\subsection{Data model storage size}
\label{subsec5}

We compared the storage requirements by measuring the total number of polygons and the total number of node required to store the processed, baseline, and the proposed data models.  The number of objects stored for both the processed and unprocessed watershed boundaries is the same $|V|$ or in our example 14,718 polygons. Because the LRG algorithm returns the highest factor of $b$ for any vertex $v$ the proposed data model also only contains $|V|$ (Table 1). The difference therefore between the  storage approaches is the number of nodes on the polygons that represent local watershed boundaries $\omega(v)$ the original graph $G$ for the baseline $\approx$ 2.16 m, processed $\approx$ 18.06 m, and the LRG algorithm $\approx$ 3.26 - 3.13 m, for $b$ (2,4,6) (Table 1).  This represents a reduction of between 81-82\% for the LRG algorithm vs the processed data model.    

The number of vertexes in the polygons that represent the catchment boundaries were originally derived from a raster grid data model representing elevation (A Digital Elevation Model). Converting a DEM to a vector polygon representation requires a large number of vertexes to represent the border of each raster grid. A number of techniques have been discussed to minimize the amount of vertexes stored on a polygon.

\begin{comment}
Use the excel file here algometrics.xlsx
\end{comment}

\subsection{Query Complexity}
\label{subsec5}
The processing time for the unprocessed data storage method requires the union of all upstream catchments on average 157 polygons and 24,550 nodes, while the processed version requires no processing time at all. The proposed method requires the output of the stitching algorithm for our example averaged between 7.46 and 6.28 polygons and 4,751 and 3,352 nodes again based on the value of $b$ Table 1.  The processed data model requires no processing and therefore has a query complexity of 1 polygon.  We used averages for these measurements instead of summing for the graph because we are most interested in queries that draw watershed boundaries on the fly for a single catchment.  The SW+LRG has a reduction in of query complexity of 95-96\% for polygons and 80-86\% for nodes on the polygon vs the unprocessed data model.

\label{subsec5}

Our results include a methodology to store and access watershed boundaries using a novel data storage technique and a readily available data set.  This data model relies on three algorithms to convert the original data into a logarithmically reduced data structure that can be merged together for any stream reach to create a geo-spatial representation surface watershed boundary $\Omega(v)$. Compared to two hypothetical data models our data models provides significant reductions in the preprocessing, data storage and query complexity necessary to create watershed boundaries.

\label{subsec5}

\begin{table}[]
\centering 
\resizebox{1 \textwidth}{!}{\begin{tabular}{|c|l|c|c|c |c|c|}
\hline
\multicolumn{2}{|c|}{}                                 & Baseline  & (LRG,SW)b=2 & (LRG,SW)b=4 & (LRG,SW)b=6 & processed \\ \hline

\multirow{2}{*}{Preprocessing}          
& \# of Polygons     & null    & 29,281      & 19,557      & 18,810      & 2,317,105       \\ \cline{2-7} 
& \# of Nodes     & null & 7,324,081   & 4,279,842   & 3,828,134   & 360,056,366   \\ \hline

\multirow{2}{*}{Storage}          & \# of Polygons     & 14,718    & 14,718      & 14,718      & 14,718      & 14,718       \\ \cline{2-7} 
                                  & \# of Nodes     & 2,162,539 & 3,260,666   & 3,261,429   & 3,136,998   & 18,063,815   \\ \hline
\multirow{2}{*}{Query Complexity} & Avg \# of Polygons & 157       & 7.46        & 6.84        & 6.28        & 1            \\ \cline{2-7} 
                                  & Avg \# of Nodes & 24,464    & 4,751       & 3,682       & 3,352       & null \\ 
\hline
\end{tabular}}
\caption{Results comparing the baseline, LRG and SW, and the processed data model}
\label{table:1}
\end{table}

\section{Discussion}
\label{subsec5}

\begin{comment}
Utility of watershed boundary delineation ( what it is used for )

How this advance could be used to move the environmental modelling field forward

Existing methods to create watershed boundaries
\end{comment}

\subsection{Integration}
\label{subsec5}
Much recent work in hydrodynamic modelling has been focused on data integration from multiple sources [11].  This work has allowed researchers to more easily access data collected from a wide variety of sources by standardizing how hydrological and weather data is stored and accessed.  Our proposed work extends these concepts by providing algorithms to process the integrated datasets. 

\subsection{Applications}
\label{subsec5}

Applications that rely on geospatial representations of watershed boundaries such as the Iowa Flood Information System (IFIS), and the Model My Watershed Tool (MMWT) are important management tools linking natural processes to geo spatial models.  IFIS for example was created to provide warnings of impending flood events.  Increases in temporal and spatial resolution of DEMs generate a need for more efficient algorithms to process and store elevation data and derived products.

There are a number of potential business applications of this model for example the Environmental Research Institute (ESRI) offers a fee for service application called Elevation Analysis as a Rest API (https://developers.arcgis.com/rest/elevation/api-reference/watershed.htm), that delineates watershed boundaries based on either the 30-meter National Elevation Dataset within the United States or the 90 meter HydroSheds database available world wide.  Unfortunately, we cannot compare our proposed data storage and access techniques to these aforementioned tools and applications as we could find no published data models.

The increasing availability of remotely sensed elevation data including the Shuttle Radar Topology Mission (SRTM) and Light Detection and Ranging (LiDAR) datasets would enable the proposed methods to be extended to other terrestrial areas. Future work will focus on how to extend the discussed data storage techniques to other watershed modelling processes including storm surge, contaminate source tracking, and groundwater flow.  

The proposed algorithms and data structures can also be applied to flow direction grids that use the D8 flow algorithm directly.  In this paper we focused on derived polygonal representations of the elevation grids.  To accomplish this each grid cell can be viewed as a vertex and the flow direction viewed as an edge.  Because the D8 method only allows one downstream edge (1 of 8 neighboring cells) the graph could be described as a tree a necessary requirement of the MNS algorithm.  We propose to test these assumptions on the flow direction grid available as part of the NHD dataset version 2, but this is outside the scope of thecurrent manuscript.

Lastly, the MNS algorithm could be applied to the existing NHD version 2 database coupled with a RDMS allowing stream routing to be accomplished using SQL.  Initial tests using the previously described server and data structure show routing calculations for the entire Delaware river completed in less then 1.5 seconds for the 14,718 reach study area. This could be very useful for mapping the location of hazardous spills, or connectivity between sources of pollutants and spills.

\section{Conclusion}

The proposed data structure provides significant reductions in the storage and query complexity costs necessary to generate watershed boundaries for the Delaware River Watershed (Table 1 and Figure 1). We believe that a number of other applications of this data model will increase the utility of existing watershed models.  These include estimates of dam removal prioritization, ecological connectivity, sums of upstream attributes such as land use and land cover, impervious surface, and population, and estimates of current stream conditions (e.g., river flooding, nutrient loading, and or stream temperature).

\newpage{}

\section{Figures}

 \begin{figure}[h!]
    \centering
    \includegraphics[scale = .9]{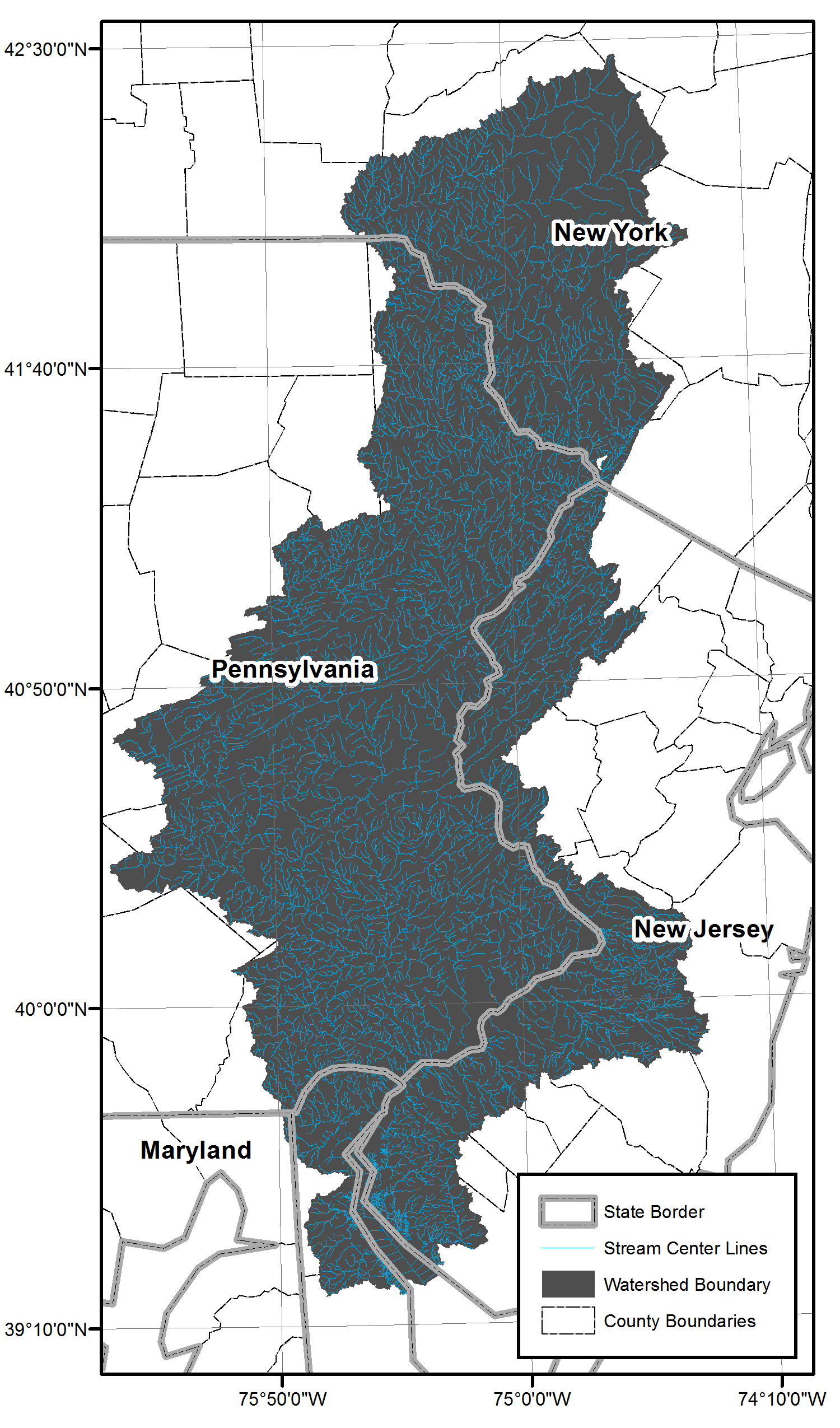}
    \caption{Overview Map of Delaware River Watershed Study Area}
     \label{Figure1:Figure 1}
\end{figure}

\newpage{}
\begin{figure}[h!] 
    \centering
    \includegraphics[scale = 0.6]{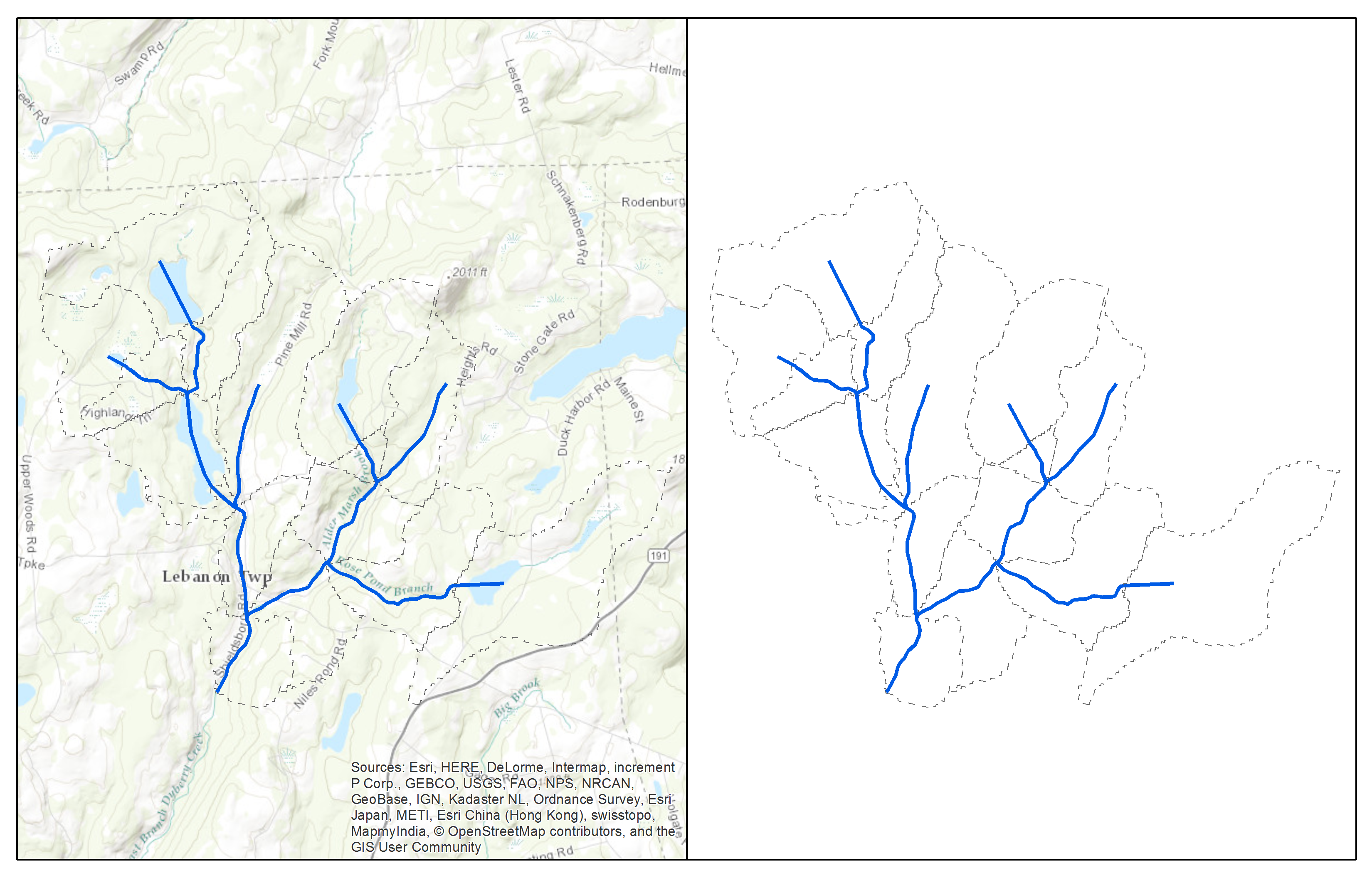}
    \caption{Map showing reach lateral watersheds}
 %   \label{Figure 1}
\end{figure}

\newpage{}
 
\begin{figure}[h!] 
    \centering
    \includegraphics[scale = 0.6]{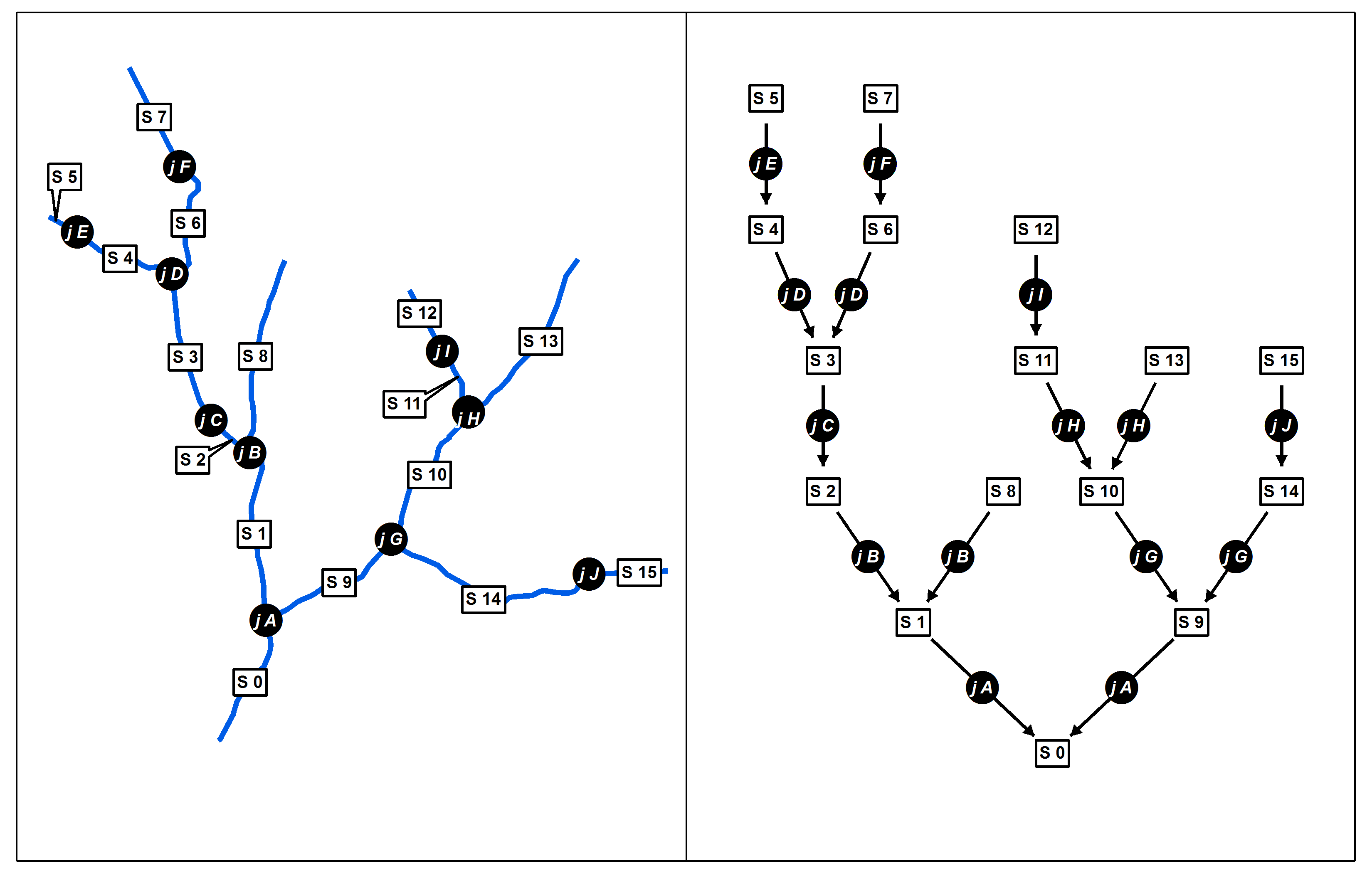}
    \caption{Relabelling stream reaches and junctions as vertices $v$ and edges $e$ in graph $G$}
    \label{Figure 2}
\end{figure} 
 \newpage{}
\begin{figure}[h!]
    \centering 
    \includegraphics[scale = 0.7]{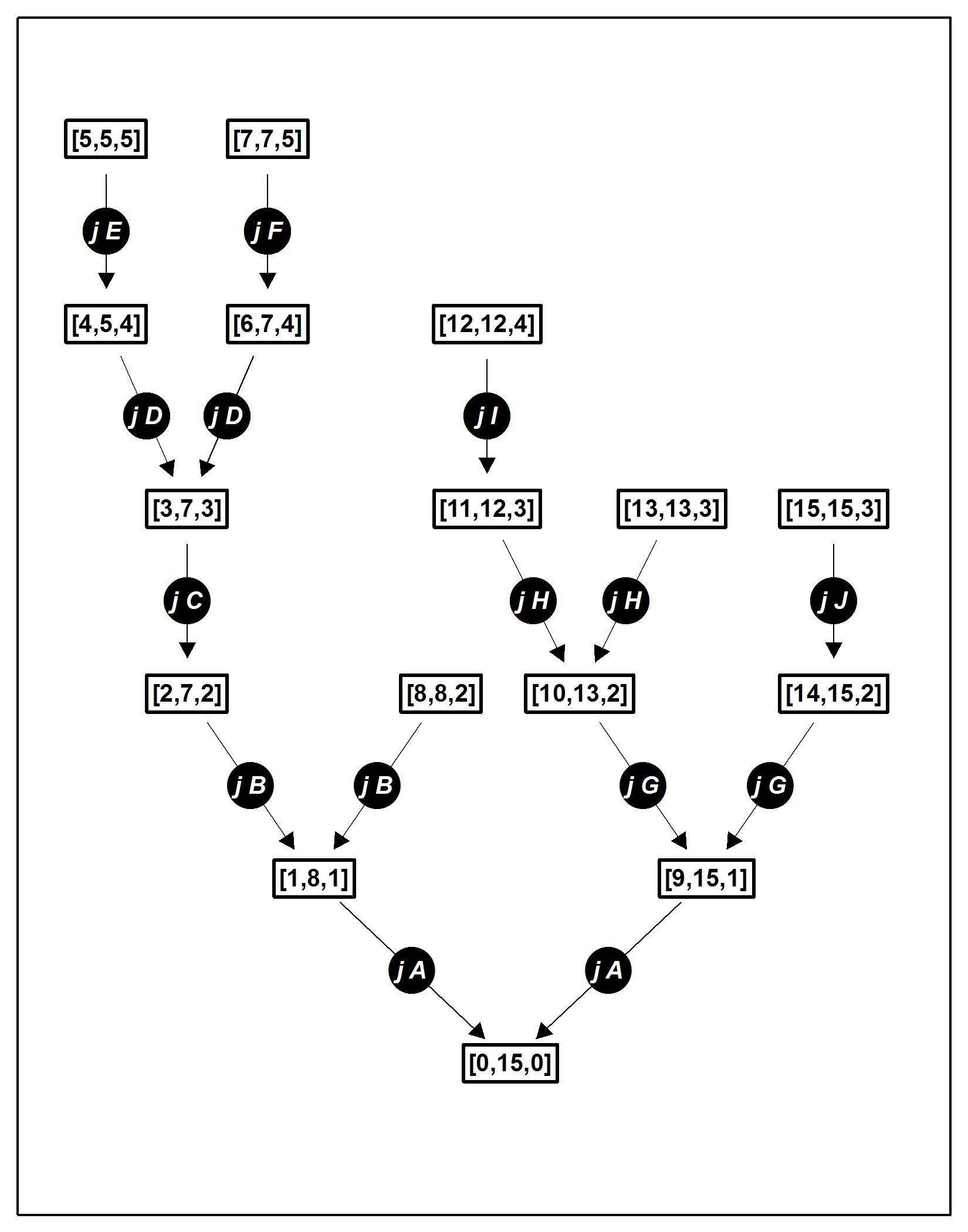}
    \caption{Results of MNS on example graph and levels upstream from the root}
 %   \label{Figure 2}
\end{figure}

\newpage{}
\begin{figure}[h!] 
    \centering
    \includegraphics[scale = .70]{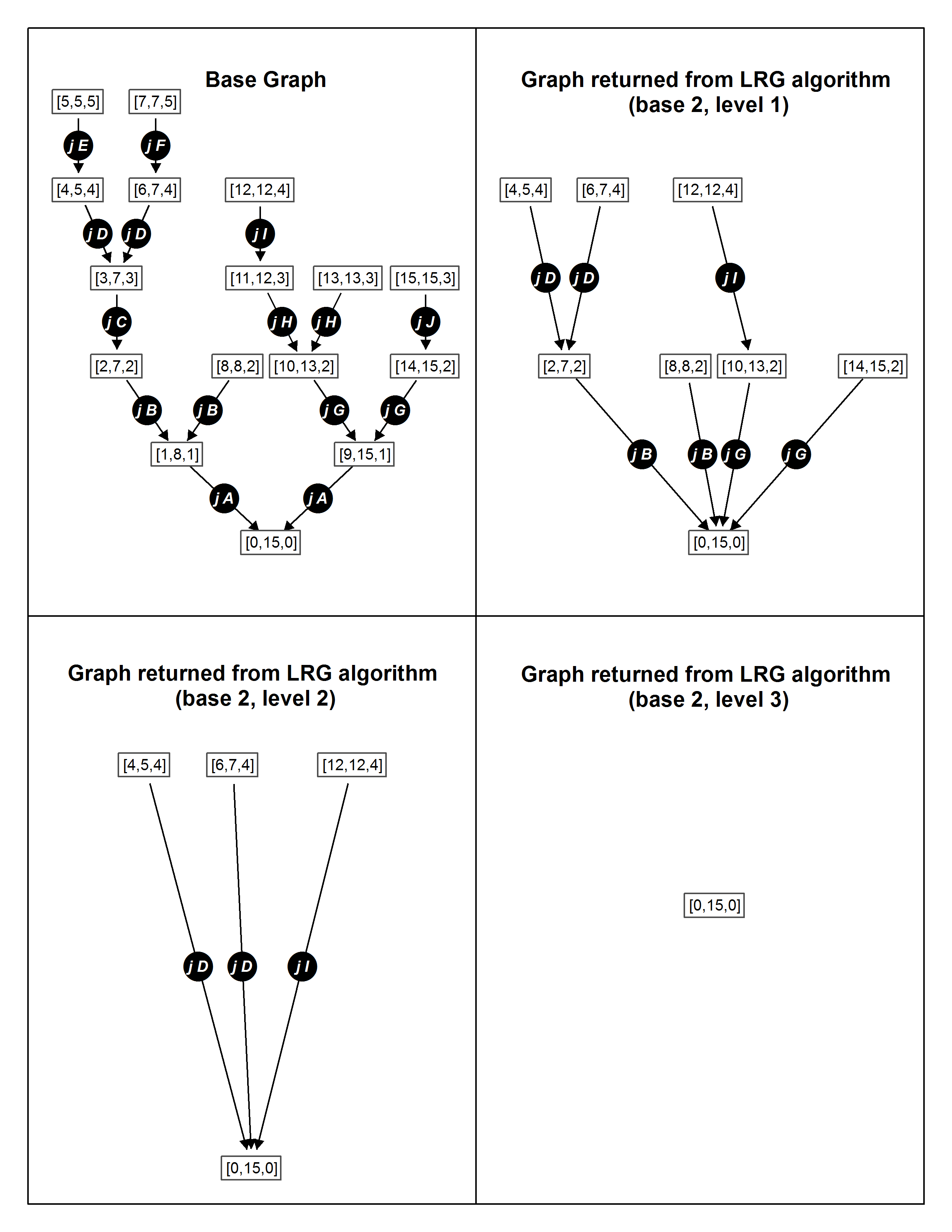}
    \caption{Results of LRG algorithm on Graph $G$, circles denote edges between vertices denoted with a box labeled showing labels returned from the MNS algorithm}
 %   \label{Figure 2}
\end{figure}

\newpage{}
\begin{figure}[h!] 
    \centering
    \includegraphics[scale = 0.71]{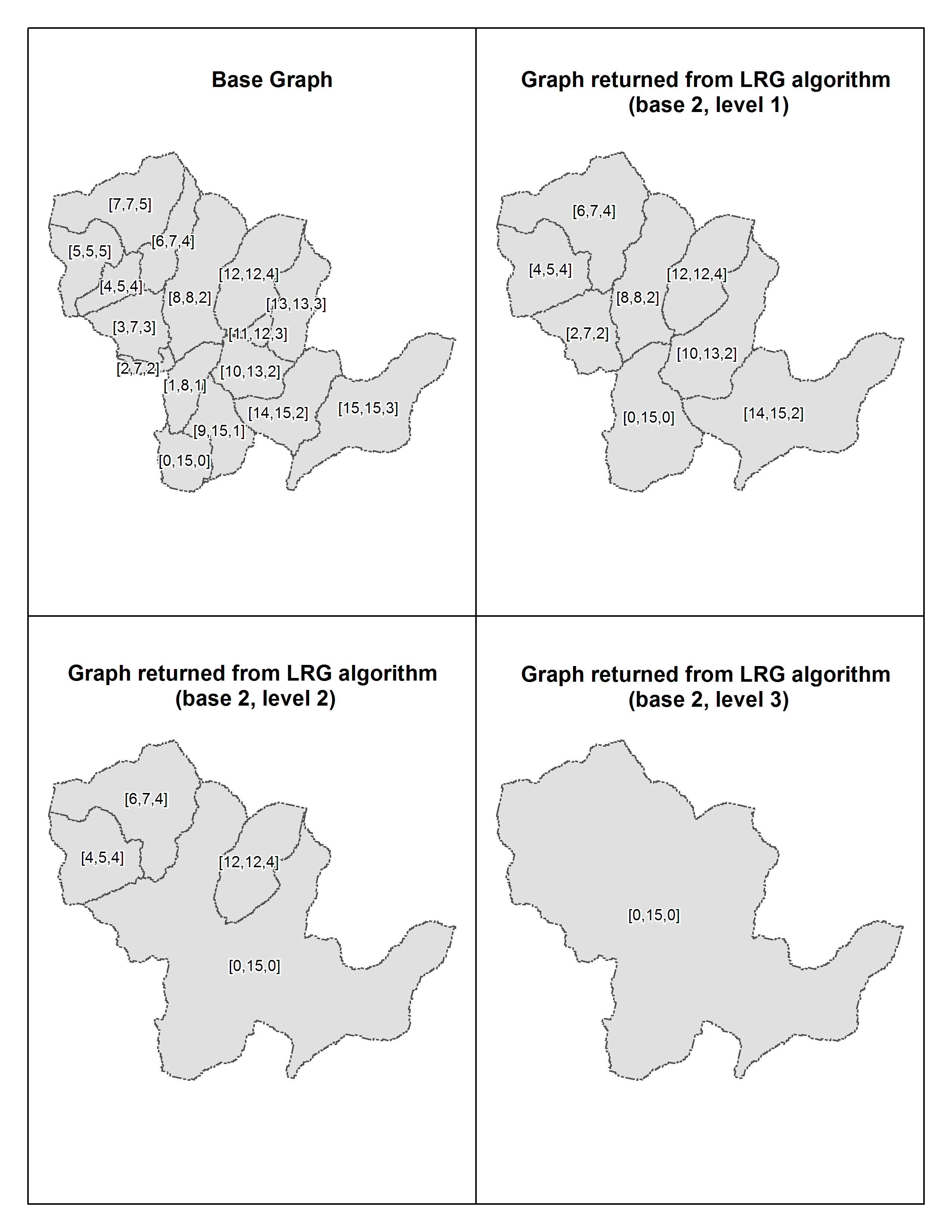}
    \caption{Catchment Boundaries for LRG graphs}
 %   \label{Figure 2}
\end{figure}

\newpage{}
\section{Acknowledgements}
We would like to acknowledge support from the William Penn foundation under grant \#39-15 'Enhanced monitoring, assessment, and data management for the Delaware River Watershed Initiative (DRWI).  The manuscript also benefited immensely from comments provided by Dr. Yusuf Osmanlioglu, Thomas Belton, and William Ryan.

\newpage{}
\section{References}

[1]  B. Lehner, C. Verdin, A. Jarvis, HydroSHEDS technical documentation.world wildlife fund US, Washington, DC.
\newline{}
[2]  L. McKay, T. Bondelid, T. Dewald, C. Johnston, R. Moore, A. Rea, NHDPlus Version 2:  User Guide(2012).
\newline{}
[3]  R. Tarjan, Amortized Computational Complexity, SIAM. J. on Algebraic and Discrete Methods 6 (2)(1985) 306–318.  doi:10.1137/0606031.URL http://epubs.siam.org/do i/abs/10.1137/0606031.
\newline{}
[4]  C. H. David, D. R. Maidment, G.-Y. Niu, Z.-L. Yang, F. Habets, V. Eijkhout, River Network Routingon the NHDPlus Dataset, J. Hydrometeor 12 (5) (2011) 913–934.  
\newline{}
[5]  F. Olivera, M. Valenzuela, R. Srinivasan, J. Choi, H. Cho, S. Koka, A. Agrawal, Arcgis-Swat: A GeodataModel  and  Gis  Interface  for  Swat1,  JAWRA  Journal  of  the  American  Water  Resources  Association42 (2) (2006) 295–309.
\newline{}
[6]  Z.-J. Liu,  D. E. Weller,  A stream network model for integrated watershed modeling,  Environmental Modeling and  Assessment 13 (2) 291–303.  doi:10.1007/s10666-007-9083-9.
\newline{}
[7] E. B. Daniel, J. V. Camp, Eugene J. LeBoeuf, Jessica R. Penrod, James P. Dobbins, Mark D. Abkowitz,Watershed Modeling and its Applications:  A State-of-the-Art Review, The Open Hydrology Journal5 (5) (2011) 26–50.
\newline{}
[8]  V. P. Singh, D. A. Woolhiser, Mathematical Modeling of Watershed Hydrology, Journal of HydrologicEngineering 7 (4) (2002) 270–292.  doi:10.1061/(ASCE)1084-0699(2002)7:4(270).
\newline{}
[9]  B. Ripley, M. Lapsley, RODBC: ODBC Database Access. URL  \newline{}https://cran.r-project.org/web/packages/RODBC/index.html.
\newline{}
[10]  J. Celko, SQL for Smartie, 4th Edition, The Morgan Kaufmann Series in Data Management Systems,Elsevier, 2011.
\newline{}
[11]  D.  P.  Ames,  J.  S.  Horsburgh,  Y.  Cao,  J.  Kadlec,  T.  Whiteaker,  D.  Valentine,  Hydrodesktop:  Webservices-based software for hydrologic data discovery, download, visualization, and analysis, Environmental Modeling and Software 37 (2) 146–156.
\newpage{}
\section{Appendices}

\appendix

\section{Section in Appendix}
\label{appendix-sec1}

\subsection{Standard Query Language SQL to create watershed for any catchment based on the results of the MNS algorithm}
\label{subsec5}

\footnotesize
\begin{algorithmic} [1]
\State{SELECT}
\State T1.$d$    
\State  ST\_UNION(t2.$\omega$) AS $\Omega$ 
\State FROM       
\State MNS($G$)            AS T1 
\State INNER JOIN 
\State MNS($G$)            AS T2 
\State ON         T2.$d$ BETWEEN T1.$d$ AND T1.$f$ 
\State WHERE      T1.$d$ = 1
\State GROUP BY   T1.$d$
\end{algorithmic} 
\normalsize

%% References


\end{document}